\documentclass[fleqn,10pt,twocolumn]{SICE_ISCS}
\usepackage{times} 
\usepackage{amsmath} 
\usepackage{amssymb}  
\usepackage{enumerate}

\usepackage{amsthm}
\newtheorem{myProblem}{\noindent\hspace{1em}\bf Problem}
\newtheorem{lemma}{\noindent\hspace{1em}\bf Lemma}

\DeclareMathOperator{\sign}{sign}
\DeclareMathOperator{\diag}{diag}

\theoremstyle{definition}
\newtheorem{definition}{\noindent\hspace{1em}\bf Definition}

\usepackage{color}
\usepackage{url}

\definecolor{forestgreen}{rgb}{0.33,0.61,0.34}
\definecolor{deepmagenta}{rgb}{0.8, 0.0, 0.8}
\definecolor{harvardcrimson}{rgb}{0.79, 0.0, 0.09}

\title{Deep unfolding-based output feedback control design for linear systems with input saturation}

\author{Koki Kobayashi${}^{1\dagger}$, Masaki Ogura${}^{2}$,
Taisuke Kobayashi${}^{1}$, and Kenji Sugimoto${}^{1}$}
\speaker{Koki Kobayashi}

\affils{${}^{1}$Division of Information Science, Nara Institute of Science and Technology, Nara, Japan\\
(Tel: +81-743-72-5356; E-mail:\{kobayashi.koki.kg9, kobayashi, kenji\}@is.naist.jp)\\
${}^{2}$Graduate School of Information Science and Technology, Osaka University, Osaka, Japan\\
(Tel: +81-6-6879-4356; E-mail: m-ogura@ist.osaka-u.ac.jp)
}
\abstract{%
In this paper, we propose a deep unfolding-based framework for the output feedback control of  systems with input saturation. Although saturation commonly arises in several practical control systems, there is still a scarce of effective design methodologies that can directly deal with the severe non-linearity of the saturation operator. In this paper, we aim to design an anti-windup controller for enlarging the region of stability of the closed-loop system by learning from the numerical simulations of the closed-loop system. The data-driven framework we propose in this paper is based on a deep-learning technique called Neural Ordinary Differential Equations. Within our framework, we first obtain a candidate controller by using the deep-learning technique, which is then tested by the existing theoretical results already established in the literature, thereby avoiding the computational challenge in the conventional design methodologies as well as theoretically guaranteeing the performance of the system. Our numerical simulation shows that the proposed framework can significantly outperform a conventional design methodology based on linear matrix inequalities.}
\keywords{%
input saturation, anti-windup, stabilization, data-driven control, deep unfolding
}

\begin{document}

\maketitle

\section{Introduction}

Saturation is often unavoidable in various control problems of practical interest including, but not limited to, spacecraft attitude control~\cite{saturation2011Asakawa}, 
vibration suppression of the mechanical equipment~\cite{saturation2013Shinohara}, and 
multi-agent systems~\cite{saturation2015Takaba}. 
It is well known that, when a conventional dynamic controller is applied to linear systems with input saturation, 
the overshoot phenomenon frequently occurs and, therefore, deteriorates system performance.
In order to address this windup phenomenon, 
an anti-windup compensator is practically used~\cite{Antiwindup2009Tarbouriech} because 
the anti-windup compensator suppresses windup by acting to stop integration 
operation of the controller when input saturation occurs. 

There are mainly two design methods for anti-windup compensators~\cite{Antiwindup2009Tarbouriech}. 
The one approach is to design the compensator to guarantee global
stability~\cite{Antiwindup1996Marcopoli,Antiwindup2001Mulder}. 
Another approach is to guarantees local stability
~\cite{Antiwindup2005Silva,antiwindup2002cao,AWC2000Kiyama,AWC2006Ichihara}. 
The design methods to guarantee global stability {often}
require the stability of the open-loop system. 
For this reason, for unstable plants, it is more practical to consider local stability. 
In this context, the estimation of the domain of attraction is an important issue. 
The domain of attraction is estimated by use of small gain theorem, Popov criterion, and circle criterion. However, it is often recognized in the literature
that the domain of attraction estimated by these methods can be  conservative~\cite{antiwindup2002cao}. 
In order to overcome this difficulty, the authors in \cite{Saturation2001Hu} propose
 a less conservative analysis approach for estimating
the domain of attraction of the systems with input saturation. 
The idea is to formulate the analysis problem into a constrained optimization problem 
with constraints given by a set of LMIs (linear matrix inequalities).

In~\cite{antiwindup2002cao}, anti-windup compensators are designed by use of 
analysis problem~\cite{Saturation2001Hu}.
However, the conditions for computing the anti-windup gains are in terms of bilinear matrix inequalities, which are difficult to solve. 
In order to overcome this difficulty, the authors in~\cite{antiwindup2002cao} propose an iterative algorithm 
based on linear matrix inequalities to solve the design problem.
It is well known that, in general, this kind of approach leads only to local optimal solutions and, furthermore, is very sensitive to the initialization process~\cite{AWdiscreteTime2001Silva}.
In~\cite{Antiwindup2005Silva}, 
the closed-loop system obtained from the dynamic feedback controller with anti-windup 
compensator can be modeled by a linear system with a deadzone nonlinearity.
A modified sector condition is then used to obtain stability conditions based on quadratic Lyapunov functions.
In this approach, the solution is direct and independent of any initialization.
These LMI-based methodologies~\cite{antiwindup2002cao,Antiwindup2005Silva} use the dynamic
controller designed in the absence of input saturation. 

In this paper, we propose a numerical method for designing a dynamic feedback controller with an anti-windup compensator 
for continuous-time systems by \emph{deep unfolding}~\cite{DeepUnfolding2019Balatsoukas}, a machine learning technique for accelerating 
the convergence of discrete iterative algorithms 
by tuning their parameters with stochastic gradient descent.
This technique allows us to simultaneously design the dynamic controller and the anti-windup compensator for a larger domain of attraction, which are separately designed in the literature~\cite{antiwindup2002cao,AWdiscreteTime2001Silva}. 

Unlike in the works~\cite{TISTA2019Ito,AverageConsensus2020Kishida} where a conventional deep unfolding is employed, the dynamical system that we deal with in this paper is operated in the \emph{continuous}-time. This difficulty poses a fundamental challenge because the feedforward signal-flow graph obtained by unfolding the closed-loop system necessarily has \emph{infinitely many} layers labeled by real numbers. 
In order to overcome this difficulty, in this paper we use Neural Neural Ordinary Differential Equations (Neural ODE)~\cite{NeuralODE2018Chen} for tuning the feedforward signal-flow graph to learn the parameters of the controllers. 
We numerically confirm that our approach can enlarge the domain of attraction of the closed-loop system.


\section{Linear systems with input saturation}

In this section, we give a brief overview of the stability analysis~\cite{Saturation2001Hu} of continuous-time linear systems with input saturation as well as methodologies for designing 
dynamic feedback controller with an anti-windup compensator. 

\subsection{System description}
Let us consider the {following} linear {time-invariant} system with actuator saturation: 
\begin{equation} \label{for:LTISystem:}
\begin{aligned}
    \frac{dx_p}{dt} & = A_px_p+B_p\sigma(u), 
    \\
    y & = C_px_p, 
\end{aligned}
\end{equation}
where $x_p\in \mathbb{R}^n$ is the state, $u\in \mathbb{R}^m$ is the control input,
$y\in \mathbb{R}^l$ is the measured output, and $A_p$, $B_p$, and $C_p$ are real-valued matrices
of appropriate dimensions. {Furthermore, the} function $\sigma \colon \mathbb{R}^m \to \mathbb{R}^m$ denotes the standard vector-valued saturation function and is defined by 
\begin{equation} \label{for:sigma_u}
    \sigma(u)_i = \sign(u_i)\min\{1, |u_i| \}
\end{equation}
for all $u\in \mathbb{R}^m$ and $i=1, \dotsc, m$.

In absence of the saturation in the system~\eqref{for:LTISystem:}, it is standard to apply a linear and dynamic controller of the form
\begin{equation}\label{for:DynamicController:}
\begin{aligned}
\frac{dx_c}{dt} &= A_cx_c+B_c y, 
\\  
u &= C_c x_c + D_c y. 
\end{aligned}
\end{equation}
where $x_c \in \mathbb{R}^{n_c}$ is the controller state. However, it is recognized in the literature that the standard controller~\eqref{for:DynamicController:} does not necessarily mitigate the deterioration of the control performance introduced by the saturation. 
A common approach to address this problem is to introduce an anti-windup compensation term~\cite{AntiWindup1999Teel} to the dynamic controller~\eqref{for:DynamicController:} as 
\begin{equation}\label{for:System:AWC}
\begin{aligned}
    \frac{dx_c}{dt} & = A_c x_c + B_cy + E_c(\sigma(u)-u), \\
    u &= C_c x_c + D_c y,
    \\
    x_c(0) &= 0, 
\end{aligned}    
\end{equation}
where $E_c \in \mathbb{R}^{n_c \times m}$ is the anti-windup gain.
The compensator~\eqref{for:System:AWC} behaves to hinder the integral operation of the 
dynamic controller when input saturation occurs.

When the anti-windup controller~\eqref{for:System:AWC} is applied to the system~\eqref{for:LTISystem:}, we obtain the following closed-loop system: 
\begin{equation}\label{for:System:}
    \begin{aligned}
   \frac{dx}{dt} &= Ax + B\left(\sigma(u)-u\right), 
   \\
   u & = Fx, 
\end{aligned}
\end{equation}
where
\begin{equation*}
x = \begin{bmatrix} x_p \\ x_c \end{bmatrix}
\end{equation*}
denotes the joint state of the closed-loop system and the matrices $A$, $B$, and $F$ are defined by 
\begin{equation*}
    \begin{aligned}
A &= \begin{bmatrix} A_p + B_pD_cC_p & B_pC_c\\
B_cC_p & A_c \end{bmatrix}, 
\\
B &= \begin{bmatrix} B_p \\ E_c \end{bmatrix}, 
\\F &= \begin{bmatrix} D_c C_p & C_c\end{bmatrix}, 
\end{aligned}
\end{equation*}
respectively. We can further rewrite the system~\eqref{for:System:} in a compact form as 
\begin{equation}\label{for:System:ODE}
    \frac{dx}{dt} = \left(A-BF\right)x+B\sigma(Fx).
\end{equation}

\subsection{Contractive invariance}

In this subsection, we describe the stability analysis~\cite{Saturation2001Hu} of the system \eqref{for:System:ODE}. We start with the following definition. 
\begin{definition}[Domain of attraction of the origin]
For $x(0)=x_0\in \mathbb{R}^{n+n_c}$, denote the state trajectory of 
the system \eqref{for:System:ODE} as $\varphi (t,x_0)$.
The domain of attraction of the origin is 
$S := \{x_0 \in \mathbb{R}^{n+n_c} : \lim_{t \to \infty} \varphi(t,x_0)=0\}$.

\end{definition}

\begin{definition}[Contractively invariant ellipsoid]
For a positive definite matrix~$P\in \mathbb{R}^{(n+n_c)\times(n+n_c)}$, define the function
\begin{equation*}
    V(x)=x^{\top}Px. 
\end{equation*}
We say that the set 
\begin{equation*}
    \Omega(P,\rho)=\left\{\tilde{x}\in\mathbb{R}^{n+n_c}:\tilde{x}^{\top}P\tilde{x}\leq\rho\right\}
\end{equation*}
is \emph{contractively invariant} with respect to the system~\eqref{for:System:ODE} if 
\begin{equation*}
    \frac{d}{dt}V(x(t)) < 0
\end{equation*}
for any trajectory of the system~\eqref{for:System:ODE}. 
\end{definition}



To state a criteria~\cite{Saturation2001Hu} for analyzing the contractive invariance of an ellipsoid, we introduce the following notations. 
For  a vector~$\nu \in \mathbb{R}^m$ and matrices~$F, H\in \mathbb{R}^{m\times (n+n_c)}$, define the matrix $M(\nu, F, H)$ by 
\begin{equation*}
    M(\nu, F, H) = (\diag \nu)F + (I - \diag \nu)H,
\end{equation*}
where $I$ denotes the identity matrix of an appropriate dimension and $\diag \nu$ denotes the diagonal matrix having the elements $\nu_1$, \dots, $\nu_m$ as its diagonal elements. Also, define the set 
\begin{equation*}
    \mathcal{V} = \{0, 1\}^m, 
\end{equation*}
which consists of 
$2^m$ vectors. 
The following lemma is given in~\cite{Saturation2001Hu}. 

\begin{lemma}[{~\cite{Saturation2001Hu}}]\label{lemma}
Let $P\in \mathbb{R}^{(n+n_c)\times (n+n_c)}$ be a positive definite matrix and let $\rho > 0$. 
The set $\Omega(P,\rho)$ is a contractively invariant set 
of the system~\eqref{for:System:ODE} if 
there exists $H\in\mathbb{R}^{m\times(n+n_c)}$
such that
\begin{align*}
&(A-BF+BM(\nu,F,H))^{\top}P \\
&+P(A-BF+BM(\nu,F,H))<0
\end{align*}
for all $\nu \in \mathcal V$ and
\begin{equation*}
\lVert Hx\rVert_\infty \leq 1 
\end{equation*}
for all $x\in \Omega(P,\rho)$, where $\lVert \cdot\rVert_\infty$ denotes the $\ell_\infty$-norm of a vector. 
\end{lemma}

Lemma~\ref{lemma} provides a condition for checking if a given ellipsoid is contractively invariant. As a scalar measure for evaluating the size of ellipsoids, we adopt the methodology developed in~\cite{Saturation2001Hu}, where the authors measure the size of ellipsoids by using a \emph{shape reference set}. Let $\mathcal{X}_R$ be a subset of $\mathbb{R}^{n+n_c}$ containing the origin. Then, for a set~$S \subset \mathbb{R}^{n+n_c} $, we define the quantity 
\begin{equation*}
    \alpha_R(\mathcal{S})=\text{sup}\{\alpha >0:\alpha \mathcal{X}_R \subset \mathcal S\}, 
\end{equation*}
which gives us a measure for the size of the set $\mathcal S$. 
Although the authors in~\cite{Saturation2001Hu} allow the shape reference set~$\mathcal{X}_R$ to be either an ellipsoid or a polyhedron, in this paper we focus on the case where the set is the polyhedron given as 
\begin{equation*}
   \mathcal{X}_R = \text{cov}\{\tilde{x}_1,\tilde{x}_2, \dotsc, \tilde{x}_l\}
\end{equation*}
by the set of pre-determined vectors~$\tilde{x}_1$, $\tilde{x}_2$, \dots, $\tilde{x}_l$ in $\mathbb{R}^{n+n_x}$. Then, by using Lemma~\ref{lemma}, one can formulate the problem of finding the maximum contractively invariant set measured by the shape reference set~$\mathcal X_R$ as the following optimization problem~{\cite{Saturation2001Hu}}: 
\begin{flalign}
&\underset{P>0, H, \rho, \alpha}{\text{maximize}}&& \alpha&& && && && && \nonumber \\
&\text{subject to} && 
\alpha \mathcal{X}_R \in \Omega(P,\rho),  \nonumber \\
& && 
\left(A-BF+BM(\nu, F, H)\right)^{\top}P \nonumber\\
& &&+P\left(A-BF+BM(\nu, F, H)\right) < 0,  \nonumber\\
& &&
\lVert Hx \rVert_\infty \leq 1,\ \mbox{for all } x  \in \Omega(P, \rho).
\label{for:estimateDoA}
\end{flalign}

It is not impossible to use the optimization problem ~\eqref{for:estimateDoA} to optimally design the parameters~$A_c$, $B_c$, $C_c$, $D_c$, and~$E_c$ in the controller~\eqref{for:System:AWC} for maximizing the contractively invariant ellipsoid. However, the optimization problem reduces to  a set of \emph{bilinear} matrix inequalities, which is not easy to solve in practice~\cite{BMI1995Toker}. Due to this difficulty, 
the conventional method~\cite{Antiwindup2005Silva} first fixes the parameters~$A_c$, $B_c$, $C_c$, and $D_c$ in such a way that the closed-loop system is stabilized in absence of the input saturation and, then, design the
anti-windup gain~$E_c$ separately. This implies that the parameters $A_c$, $B_c$, $C_c$, and $D_c$ are not necessarily designed for enlarging the contractively invariant ellipsoid. In order to fill in this gap, in the next section we propose a novel deep unfolding-based approach for designing all the parameters simultaneously for enlarging the contractively invariant ellipsoid. 

\section{Proposed method}

In this section, we describe our framework of deep unfolding-based design of the controller for maximizing the contractively invariant ellipsoid of the closed-loop system~\eqref{for:System:ODE}. 

\subsection{Problem formulation}

As mentioned in the previous section, it is difficult to effectively design a dynamic  anti-windup controller by using the optimization problem~\eqref{for:estimateDoA}. In order to avoid this difficulty, in this paper we propose the following approach we first design a \emph{candidate} controller by using a date-driven, deep learning-based approach described below. 
We then \emph{test} the validity of the candidate controller by using the \emph{linear} matrix inequalities~\eqref{for:estimateDoA}.  

In order to obtain a candidate controller, we propose that the following  problem is solved:

\begin{myProblem}\label{Problem:1}
Let $T$ be a positive number and $\mathcal{C} \subset \mathbb{R}^{n+n_c}$ be a set. 
Assume that the initial state~$x_0$ of the {closed-loop} system~\eqref{for:System:ODE} is a random variable
uniformly distributed on a set~$\mathcal{C} \subset \mathbb{R}^{n+n_c}$.
Find the matrices $A_c$, $B_c$, $C_c$, $D_c$ and $E_c$ such that 
\begin{equation}\label{eq:myu}
    \mu = \mathbb{E}\left[\int^{T}_{0}\|x(t)\|^2dt\right]
\end{equation}
is minimized, where $\mathbb{E}[\cdot]$ denotes the mathematical expectation with respect to the distribution of the initial state~$x_0$. 
\end{myProblem}

In order to describe our motivation for introducing Problem~\ref{Problem:1}, we below show how we can intuitively relate Problem~\ref{Problem:1} and 
the controller design problem. 
Suppose that the {region}~$\mathcal{C}$ satisfies 
$\mathcal{C}\subset\Omega(P,\rho)$ for a set of parameters $(A_c, B_c, C_c, D_c, E_c)$. 
Then, the state trajectory of the system \eqref{for:System:ODE} for $x(0)\in\mathcal{C}$ satisfies 
$\lim_{t\to\infty}x(t)=0$.
Therefore, when $T$ is  sufficiently large, 
we can expect that the set of the gains gives a good solution to Problem~\ref{Problem:1}. 
Conversely, if the solution of Problem~\ref{Problem:1} for sufficiently large~$T$ makes the value of the objective function $\mu$ sufficiently small, i.e., when 
$x(t)$ is sufficiently small {at the $T$} 
for the initial state $x(0) \in \mathcal{C}$, then we can expect that 
the ellipsoid $\Omega(P,\rho)$ obtained in this case is larger than the region~$\mathcal{C}$.


\subsection{Deep unfolding-based design}

Although one can expect that the solution of Problem~\ref{Problem:1} could allow us to design a feedback controller for enlarged invariant set, it is not straightforward to directly minimize the quantity  $\mu$ given in \eqref{eq:myu}. In this paper, we show that we can efficiently perform the minimization by using the technique called deep unfolding. 

In the framework of deep unfolding, adjust parameters using deep learning
techniques, the signal-flow of the iterative algorithm is 
unfolded, and the network with the structure of this 
signal-flow graph is created.
Therefore, in this paper, we unfold the signal-flow of the system~\eqref{for:System:ODE} to obtain a signal-flow graph. For training this network, we propose that ODE-Net~\cite{NeuralODE2018Chen} is used.
In ODE-Net, it is allowed that the structure of the network is represented 
by an ordinary \emph{differential} equation $\dot x(t) = f(x(t),t,\theta)$, with $\theta$ being a trainable parameter.
Therefore, we construct a network with the structure of the state equation~\eqref{for:StateEquation:ODE} as 
\begin{align}
   \frac{dx(t)}{dt} &=f(x(t),t,A_c,B_c,C_c,D_c,E_c) \notag\\
   &=(A-BF)x+B\sigma(Fx) \label{for:StateEquation:ODE}
\end{align}
for using ODE-Net to learn the gain of a dynamic controller with an anti-windup compensator.

We employ the technique of incremental learning~\cite{TISTA2019Ito} to efficiently train the network.
We specifically determine the number of incremental learning iterations $N$ and let {$t_k = kT/N$} for $k=1$, \dots, $N$. We then calculate the candidate controller parameter sets $(A_{c,k},B_{c,k},C_{c,k},D_{c,k},E_{c,k})$ 
to minimize the loss function
\begin{equation*}
    L(0, t_k)= \mathbb{E}\left[ \int^{t_k}_{0}\|x(t)\|^2dt \right]
\end{equation*}
at each step.
We then solve an optimization problem \eqref{for:estimateDoA} for 
$A_{c,k}$, $B_{c,k}$, $C_{c,k}$, $D_{c,k}$, and $E_{c,k}$ 
at each step of the incremental training, and use
the controller with the largest ellipsoid. 

\begin{figure}[tb]
    \centering
    \includegraphics[width=80mm]{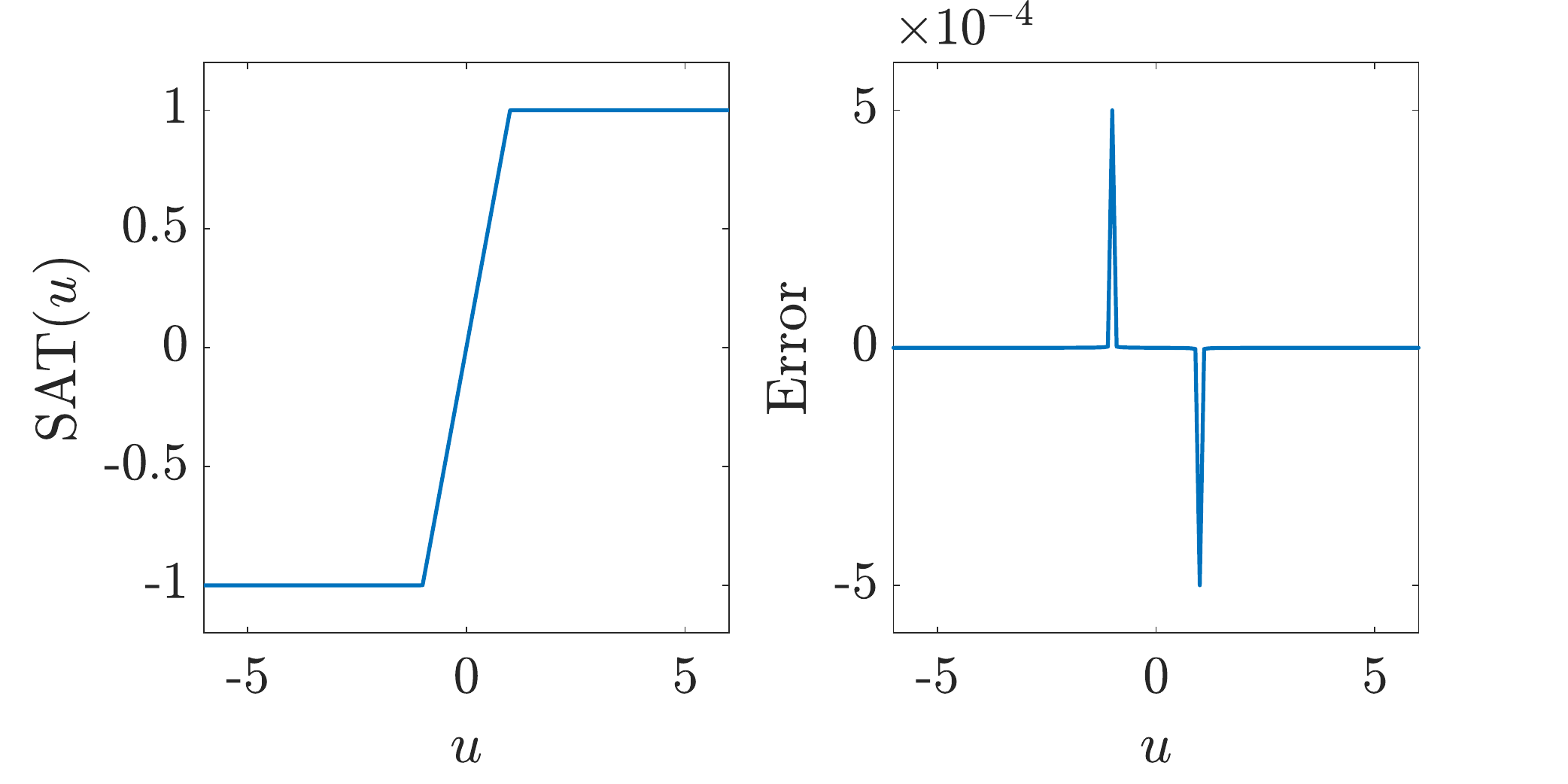}
    \caption{approximate function (left) evaluated for $\zeta=1.0\times10^{-6}$, approximation
error (right)}
    \label{fig:SATAPPFunc}
\end{figure}

To effectively enlarge the ellipsoid, we set the region~$\mathcal C$ as 
\begin{equation*}
    \mathcal{C}=(\beta\Omega(P_{0},\rho_{0}))\backslash\Omega(P_{0},\rho_{0})
\end{equation*}
for some parameter~$\beta>1$, and train the network by using the initial states outside the obtained region~$\Omega(P_{0},\rho_{0})$. 
In order to implement the process described above, one may use \texttt{DiffEqFlux.jl}~\cite{Rackauckas2019DiffEq} together with  \texttt{DifferentialEquations.jl}~\cite{Diffeq2017Chris}. 
We remark that, because these tools are only applicable to dynamical systems with smooth vector fields, it is necessary to to approximate the non-smooth saturation function~\eqref{for:sigma_u} by a smooth function. For example, one can use the approximation of the form (see, e.g., \cite{Avvakumov2000Sigma})
\begin{equation}
    \sigma(u) \approx \frac{\sqrt{\zeta+\left(u+1\right)^2}}{2}-\frac{\sqrt{\zeta+\left(u-1\right)^2}}{2}. 
\label{for:sigma:app}
\end{equation}
In Fig.~\ref{fig:SATAPPFunc}, we present a comparison of the original saturation mapping and its approximation~\eqref{for:sigma:app} for the value $\zeta=1.0\times10^{-6}$, in which a small approximation error is achieved. 

In summary, we below present the design algorithm that we propose in this paper: 
\begin{enumerate}[Step 1)]
    \item Let a  reference set $\mathcal{X}_R$ and initial gains $A_{c,0}$, $B_{c,0}$, $C_{c,0}$, $D_{c,0}$ be given. Let $E_{c,0}=0$ and solve the optimization problem~\eqref{for:estimateDoA}.
    Denote the solutions of the optimization problem as $\alpha_0$ and $P_0$.
    Set 
    \begin{equation*}
    \mathcal{C}=(\beta\Omega(P_0,\rho_0))\backslash\Omega(P_0,\rho_0)    
    \end{equation*}
    and $\alpha_{\max}=\alpha_0$.
    \item Set $k=1$. 
    \item 
          Generate $J$ initial states $\tilde{x}(0)=\tilde{x}_{k,1}$, \dots, $\tilde{x}_{k,J}$ randomly and uniformly from region~$\mathcal{C}$.\label{step:lm}
    \item Use ODE-Net to train the network under the loss function $L(0,t_{k})$ and find a set of controller parameters. 
    \item For $A_{c,k}$, $B_{c,k}$, $C_{c,k}$, $D_{c,k}$, and $E_{c,k}$, 
          Solve the optimization problem~\eqref{for:estimateDoA}. 
              Denote the solution as $\alpha_k$. If $\alpha_{\max}<\alpha_k$, let 
        $\alpha_{\max}=\alpha_k$, $A_c=A_{c,k}$, $B_c=B_{c,k}$, $C_c=C_{c,k}$, $D_c=D_{c,k}$, and $E_c=E_{c,k}$.
    \item If $k<N$, increment $k$ and go to step (\ref{step:lm}). Else, go to step (\ref{step:mn}).
    \item $A_c$, $B_c$, $C_c$, $D_c$ and $E_c$ are controller parameters and stop. 
          \label{step:mn}
\end{enumerate}

\section{PERFORMANCE EVALUATION}

\begin{figure}[bt]
    \centering
    \includegraphics[width=80mm]{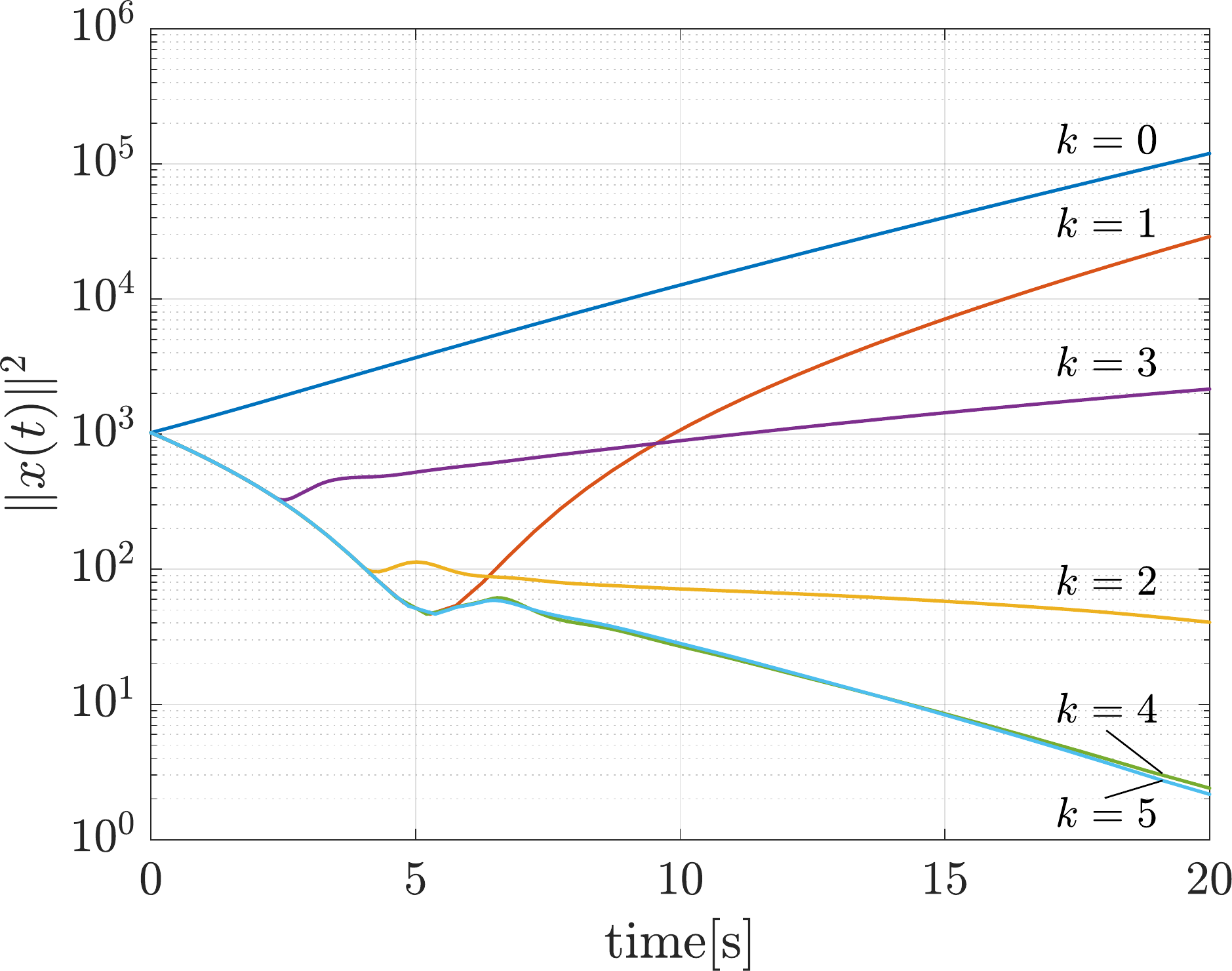}
    \caption{Norm $\|x(t)\|^2$ with the gain when $k=0,1,\cdots,5$.}
    \label{fig:AWCPlot_j}
\end{figure}
\begin{figure}[bt]
    \centering
    \includegraphics[width=80mm]{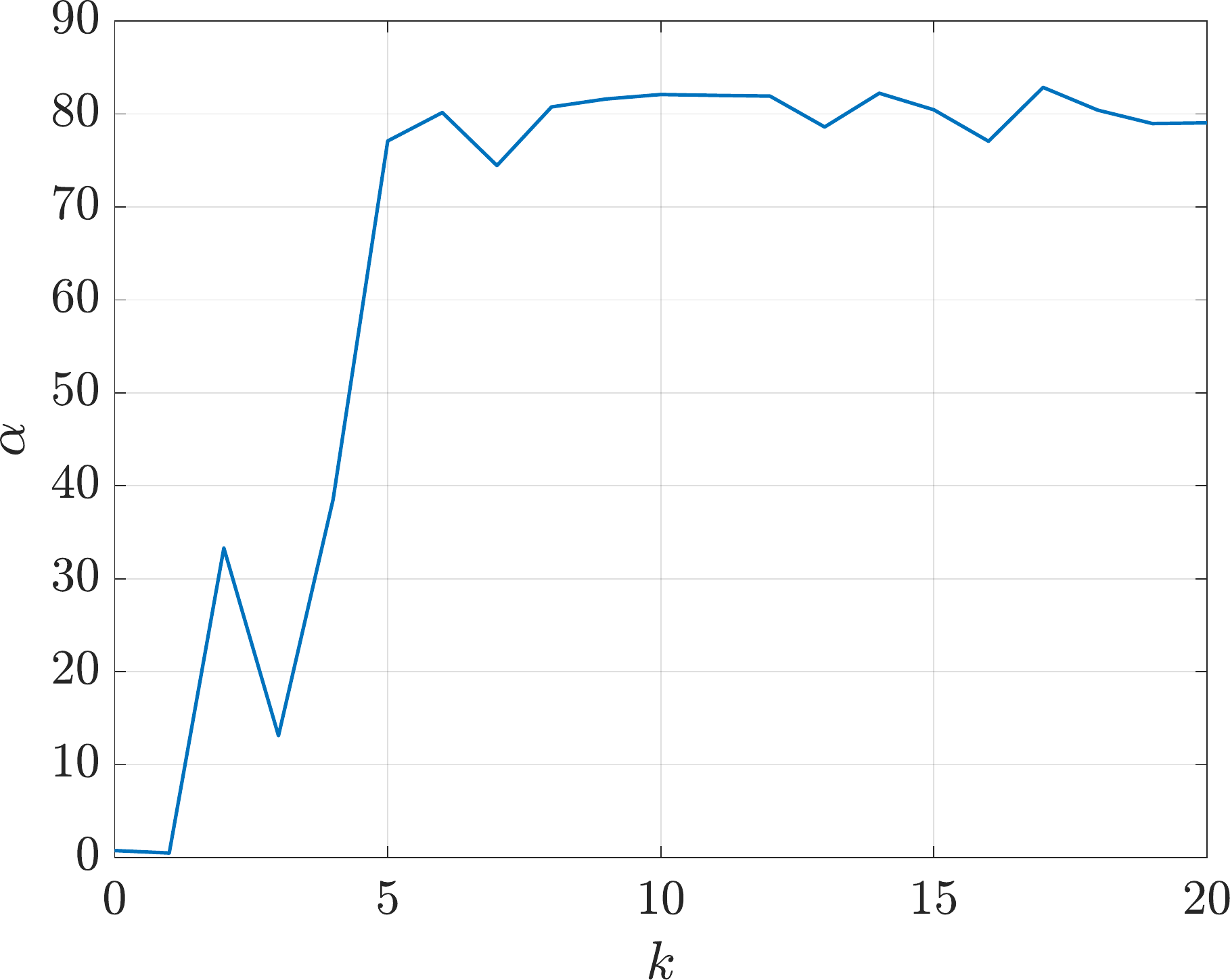}
    \caption{The value of the parameter $\alpha$ at each step of the training.}
    \label{fig:alphaPlot}
\end{figure}
\begin{figure}
    \centering
    \includegraphics[width=80mm]{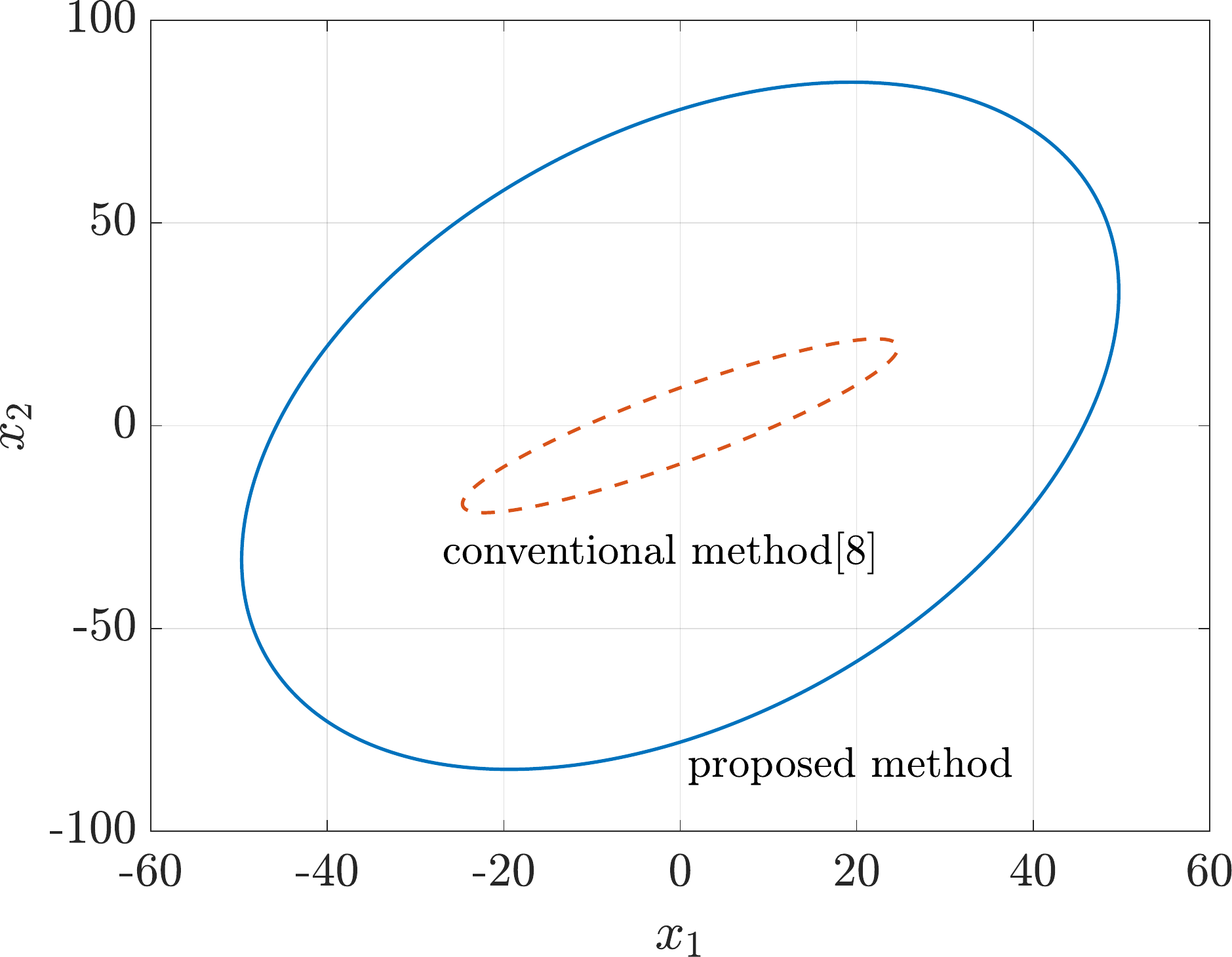}
    \caption{Ellipsoid: Proposed approach (continuous). Conventional method~\cite{antiwindup2002cao} (dashed). }
    \label{fig:PlotEllipsoid}
\end{figure}
\begin{figure}
    \centering
    \includegraphics[width=80mm]{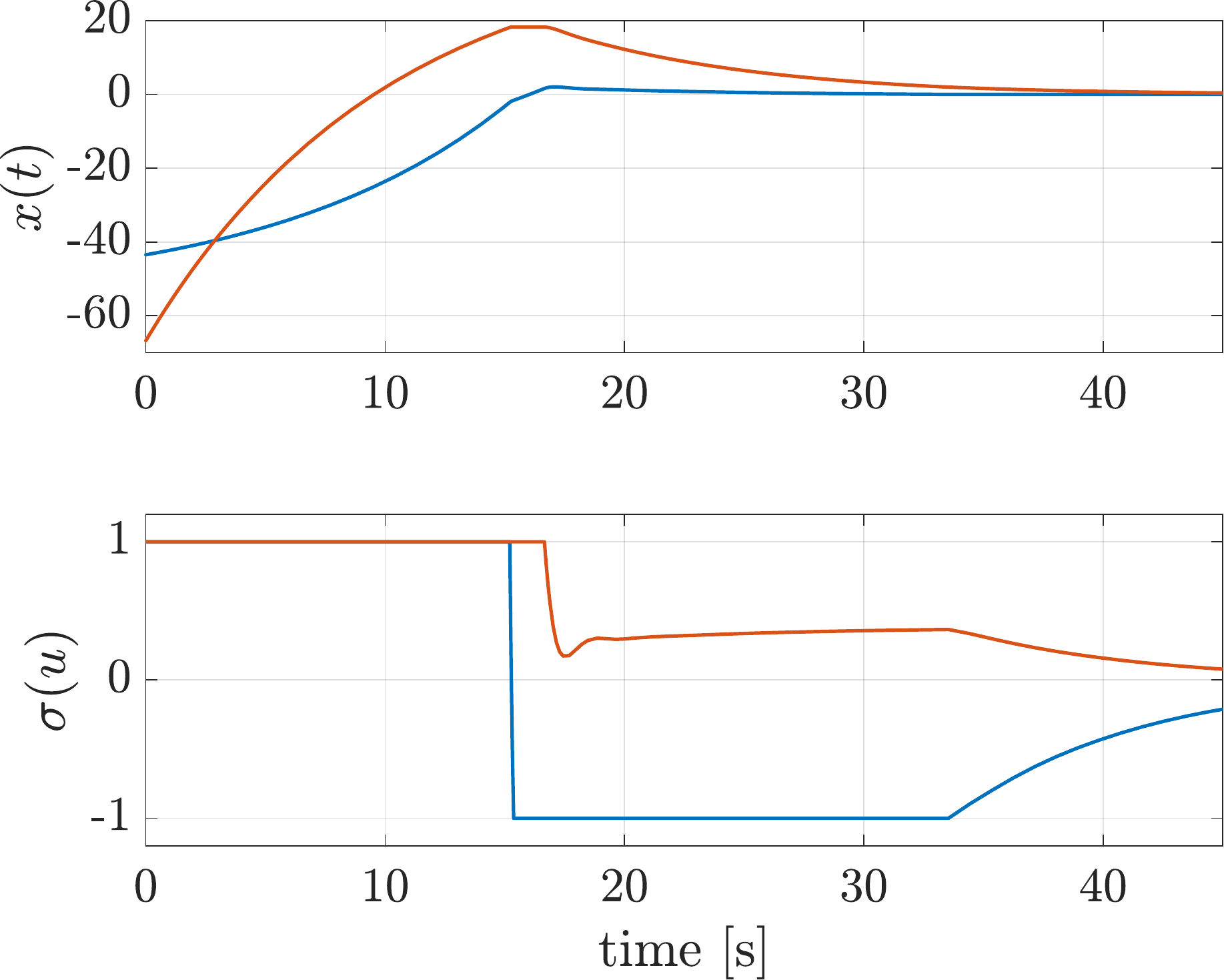}
    \caption{State response with proposed controller gain.}
    \label{fig:PlotReaction}
\end{figure}

In this section, we demonstrate the effectiveness of the proposed method based on numerical
experiments.
To demonstrate the effectiveness of our design method, let us consider the following open-loop unstable plant~\cite{Antiwindup2005Silva}: 
\begin{equation}
\begin{aligned}
   \dot{x}_p(t)&= \begin{bmatrix} 0.1 & 0 \\ 0 & -0.1\end{bmatrix}x_p(t)+
    \begin{bmatrix}1.5 & 4 \\ 1.2 & 3 \end{bmatrix}\sigma\left(u(t)\right), \\
    y(t) &= \begin{bmatrix} 1 & 0 \\ 0 & 1\end{bmatrix}. 
\end{aligned}\nonumber
\label{for:ex:system}
\end{equation}
The initial value of the dynamic controller is set to be 
\begin{align}
   \dot{x}_c(t) &= \begin{bmatrix} 0 & 0 \\ 0 & 0\end{bmatrix} x_c(t)+ \begin{bmatrix}-1 & 0 \\ 0&-1\end{bmatrix}y \nonumber\\
    u(t) &= \begin{bmatrix}0.3333 & 0 \\ 0 & -0.1 \end{bmatrix}x_c +\begin{bmatrix} -3.3333 & 0 \\ 0 & 1\end{bmatrix},  \nonumber
\end{align}
which is the same as the one used in~\cite{Antiwindup2005Silva}.  
Also, we let,
$
\mathcal{X}_R=\begin{bmatrix}x_p^{\top}(0) & x_c^{\top}(0)\end{bmatrix}^{\top}
\label{for:ex1:XR}
$, 
with $x_p(0)=\begin{bmatrix}0.6 & 0.4\end{bmatrix}^{\top}$ and 
$x_c(0) = 0$.
Also, let $T=20$, $J=10$, $N = 20$, $\beta = 10.0$.
Based on this setup, experiments were conducted in 
\texttt{DiffEqFlux.jl}~\cite{Rackauckas2019DiffEq} using Adam with 
a learning rate of $0.01$ and in 
\texttt{DifferentialEquations.jl}~\cite{Diffeq2017Chris} using Runge-Kutta $4$th order method for training.
The optimization problem \eqref{for:estimateDoA} is based on a reference set $\mathcal{X}_R$ 
to find a larger ellipsoid centerd on the origin.
Since the elements of the reference set $\mathcal{X}_R$ exist only in the first quadrant,
the initial value of step~\ref{step:lm} of the design algorithm is randomly selected form
the first and third quadrants of the region.
The norm $\|x(t)\|^2$ of the system when the initial value of the system is 
$x_p(0)=\begin{bmatrix}-25.0 & -20.0\end{bmatrix}$ for the gains at $k=0,1,\dots,5$ is shown in the Fig.~\ref{fig:AWCPlot_j}.
Fig.~\ref{fig:AWCPlot_j} shows that it is possible to find the gain that converges the state of the system to zero by applying the proposed method.
$\alpha_j$ is shown in Fig.~\ref{fig:alphaPlot}.
In Fig.~\ref{fig:alphaPlot}, $\alpha$ is increasing. 
Therefore, by solving problem~\ref{Problem:1} using an algorithm 
based on deep unfolding, 
we can find the gain to enlarge the ellipsoid.
Also, $\alpha$ is the largest when $k=17$ and $\alpha=82.858$, 
and each gain of the dynamic controller with anti-windup compensator 
was
\begin{align}
    A_c &=\begin{bmatrix} -0.9134 & 0.5888 \\ -0.4153 & -1.5813\end{bmatrix},\nonumber \\
        B_c &=\begin{bmatrix} -1.2601 & -0.2081 \\ 0.3309 & -0.6955\end{bmatrix}, \nonumber\\
    C_c &=\begin{bmatrix} -0.1852 & 0.6730 \\ 0.2955 & -0.4372 \end{bmatrix}, \nonumber\\
    D_c &= \begin{bmatrix} -3.9034 & -0.3487 \\ -0.8045 & 0.2136 \end{bmatrix}, \nonumber\\
    E_c &= \begin{bmatrix} -0.0195 & 1.5041 \\ 0.4874 & -1.3736 \end{bmatrix}. \nonumber
\end{align}
In~\cite{Antiwindup2005Silva}, when $E_c^2\leq100$, 
the parameter indicating 
the size of the ellipsoid region is set to prevent high gain, 
${\tilde{\alpha}=36.6119}$, and without the constraint, ${\tilde{\alpha}=71.72}$. 
Using the itelative LMI~\cite{antiwindup2002cao}, 
we obtained $\alpha = 40.4398$ when $E^2_c \leq 10000$, and we could not find 
a feasible solution in the absence of constraints.
These values are obtained by solving a different optimization problem than 
optimization problem \eqref{for:estimateDoA}. 
However, the results in~\cite{antiwindup2002cao} and~\cite{Antiwindup2005Silva} are shown to be 
comparable~\cite{Antiwindup2005Silva}.
The proposed method results in a larger ellipsoid region than the
conventional method~\cite{Antiwindup2005Silva,antiwindup2002cao}. 
The ellipsoid corresponding to the gain obtained by the proposed method 
and conventional method~\cite{antiwindup2002cao}.
for $x_c = 0 $ is shown in Fig.~\ref{fig:PlotEllipsoid}.
Fig.~\ref{fig:PlotEllipsoid} shows that the system designed by the proposed method is more stable 
than the system designed by the conventional method~\cite{antiwindup2002cao} in a wide range.
Fig.~\ref{fig:PlotReaction} shows a state response under 
the calculated gain. The initial condition is $x_p=\begin{bmatrix}-43.48&-66.78\end{bmatrix}^{\top}$.
This initial condition is the value on the border of the ellipsoid.
Fig.~\ref{fig:PlotReaction} shows that the values of the inputs are within the limits, and the state of the system is converged.

We verify that the proposed method is also valid 
for different reference sets.
To show its validity, let the reference set be 
\begin{equation*}
 \mathcal{X}_R = \left\{\begin{bmatrix}1\\1\end{bmatrix},\begin{bmatrix}1\\-1\end{bmatrix}\right\}.  
\end{equation*}
We will compare the initial condition $x_p(0)$, 
which we give when designing, with the same case as in the example above,
and the case where we choose from all the regions in region~$\mathcal{C}$.

Let $T=20,J=40$.
When the initial condition was selected from the first and 
third quadrants as in the previous example, 
$\alpha = 0.0034$, and when the initial condition was selected from all regions, $\alpha = 20.216$.
If the value of the reference set is changed, a controller with a larger 
ellipsoid can be designed by changing the initial condition of the system
given at the time of design.
Applying~\cite{antiwindup2002cao}, we obtain $\alpha=8.4737$.
In the case of changing the reference set, the ellipsoid of the proposed
method became larger.

\section{conclusion}
In this note, we have proposed the deep unfolding-based design method of output feed controller for continuous-timesystems with input saturation.
In the conventional method~\cite{Antiwindup2005Silva,antiwindup2002cao},
the dynamic controller is designed 
so that the closed-loop system becomes stable 
without considering input saturation, 
and the anti-windup controller is designed.
On the other hand, proposed method design a dynamic controller and 
an anti-windup compensator considering input saturation.

In the numerical examples, we designed two reference sets to 
show the effectiveness of the proposed method.
In the first numerical example, the controller designed 
with the proposed method has a larger absorption region than the 
conventional method~\cite{Antiwindup2005Silva,antiwindup2002cao}.
In the second numerical example, we changed the reference set.
The proposed method enables us to enlarge the ellipsoid 
by changing the range of the initial state according to the reference set.
Even when the reference set is changed, 
the ellipsoid is larger than that of the design 
by the conventional method.


\begin{thebibliography}{99}
\bibitem{saturation2011Asakawa}
T. Asakawa, I. Yamaguchi, Y. Hamada, T. Kasai, T. Nagashio, and T. Kida, 
``Anti-windup design for flexible spacecraft attitude control system'', 
in \textit{SICE Annual Conference 2011}, pp. 1803--1806, 2011.
\bibitem{saturation2013Shinohara}
Y. Shinohara, K. Seki, M. Iwasaki, H. Chinda, and M. Takahashi,
``Controller design for dual-stage actuator-driven load devices considering suppression of vibration due to input saturation'', 
in \textit{2013 IEEE International Conference on Mechatronics}, 
pp. 742–747, 2013.
\bibitem{saturation2015Takaba}
K. Takaba, 
``Local synchronization of linear multi-agent systems subject to input saturation'',
\textit{SICE Journal of Control, Measurement, and System Integration},
Vol. 8, No. 5, pp. 334--340, 2015.
\bibitem{Antiwindup2009Tarbouriech}  
S. Tarbouriech and M. Turner,
``Anti-windup design:  an overview of some recent advances and open problems'', 
\textit{IET Control Theory \& Applications},
Vol. 3, No. 1, pp. 1--19, 2009.
\bibitem{Antiwindup1996Marcopoli}
V. R. Marcopoli and S. M. Phillips, 
``Analysis and synthesis tools for a class of actuator-limited multivariable control systems: A linear matrix inequality approach'', 
\textit{International Journal of Robust and Nonlinear Control}, 
Vol. 6, No. 9-10, pp. 1045--1063, 1996.
\bibitem{Antiwindup2001Mulder}
E. F. Mulder, M. V. Kothare, and M. Morari, 
``Multivariable anti-windup controller synthesis using linear matrix inequalities'', 
\textit{Automatica}, Vol. 37, No. 9, pp. 1407--1416, 2001.
\bibitem{Antiwindup2005Silva}
J. M. G. da Silva and S. Tarbouriech, 
``Antiwindup design with guaranteed regions of stability: an LMI-based approach'', 
\textit{IEEE Transactions on Automatic Control}, 
Vol. 50, No. 1, pp. 106–111, 2005.
\bibitem{antiwindup2002cao}
Y. Y. Cao, Z. Lin, and D. G. Ward. 
``An antiwindup approach to enlarging domain of attraction for linear systems subject to actuator saturation'', 
\textit{IEEE Transactions on Automatic Control}, Vol. 47, No. 1, pp. 140--145, 2002.
\bibitem{AWC2000Kiyama}
T. Kiyama and T. Iwasaki, 
``On the use of multi-loop circle criterion for saturating control synthesis'', 
\textit{Systems \& Control Letters}, 
Vol. 41, No. 2, pp. 105--114, 2000.
\bibitem{AWC2006Ichihara}
H. Ichihara and E. Nobuyama, 
``Stability analysis and control design of linear systems with input saturation using matrix sum of squares relaxation'', 
\textit{IFAC Proceedings Volumes}, vol.39, no. 9, pp. 369--374, 2006.
\bibitem{Saturation2001Hu}
T. Hu, Z. Lin, B. M. Chen,
``An analysis and design method for linear systems subject to actuator saturation and disturbance'', 
\textit{Automatica},
Vol. 38, No. 2, pp. 351--359, 2002,

\bibitem{AWdiscreteTime2001Silva}
J. M. G. Da Silva and S. Tarbouriech, 
``Local stabilization of discrete-time linear systems with saturating controls: an LMI-based approach'', 
\textit{IEEE Transactions on Automatic Control}, Vol. 46, No. 1, pp. 119--125, 2001.

\bibitem{DeepUnfolding2019Balatsoukas}
A. Balatsoukas and C. Studer,
``Deep Unfolding for Communications Systems: A Survey and Some New Directions'', 
in \textit{2019 IEEE International Workshop on Signal Processing Systems}, pp. 266--271, 2019.
\bibitem{TISTA2019Ito} 
D. Ito, S. Takabe, and T. Wadayama,
``Trainable ISTA for sparse signal recovery'', 
\textit{IEEE Transactions on Signal Processing}, 
Vol. 67, No. 12, pp. 3113--3125, 2019.
\bibitem{AverageConsensus2020Kishida}
M. Kishida, M. Ogura, Y. Yoshida, and T. Wadayama, 
``Deep learning-based average consensus'', 
\textit{IEEE Access}, Vol. 8, pp. 142404--142412, 2020.
\bibitem{NeuralODE2018Chen}
R. T. Q. Chen, Y. Rubanova, J. Bettencourt, and D. K Duvenaud, 
``Neural ordinary differential equations'', 
\textit{32nd Conference on Neural Information Processing Systems}, 
pp. 6571--6583, 2018.
\bibitem{AntiWindup1999Teel}
A. R. Teel, 
``Anti-windup for exponentially unstable linear systems'',
\textit{International Journal of Robust and Nonlinear Control}, 
Vol. 9, No. 10, pp. 701--716, 1999.
\bibitem{BMI1995Toker}
O. Toker and H. Ozbay, 
``On the NP-hardness of solving bilinear matrix inequalities and simultaneous stabilization with static output feedback'', 
in \textit{1995 American Control Conference}, 
Vol. 4, pp. 2525--2526, 2005.
\bibitem{Rackauckas2019DiffEq}
C. Rackauckas, M. Innes, Y. Ma,J. Bettencourt, L. White, and V. Dixit, 
``DiffEqFlux.jl - A julia library for neural differential equations'', 
	arXiv:1902.02376, 2019.
\bibitem{Diffeq2017Chris}
C. Rackauckas and Q. Nie, 
``Differentialequations.jl – a performant and feature-rich ecosystem for solving differential equations in julia'', 
\textit{Journal of Open Research Software}, 
Vol. 5, No. 1, p. 15, 2017. 
\bibitem{Avvakumov2000Sigma} 
P. Boscariol and A. Gasparetto, 
``Optimal trajectory planning for nonlinear systems: robust and constrained solution'',
\textit{Robotica}, 
Vol. 34, No. 6, pp. 1243-1259, 2016. 
\end{thebibliography}
\end{document}